\documentstyle[12pt]{article}
\newcommand{\pbm}{\mbox{\boldmath $p$}}
\newcommand{\fbm}{\mbox{\boldmath $f$}}
\newcommand{\rbm}{\mbox{\boldmath $r$}}
\newcommand{\sigmabm}{\mbox{\boldmath $\sigma$}}
\newcommand{\xibm}{\mbox{\boldmath $\xi$}}
\newcommand{\lambdabm}{\mbox{\boldmath $\lambda$}}
\makeatletter
\def\lsim{\mathrel{\mathpalette\gl@align<}}
\def\gsim{\mathrel{\mathpalette\gl@align>}}
\def\gl@align#1#2{\lower.6ex\vbox{\baselineskip\z@skip\lineskip\z@
    \ialign{$\m@th#1\hfil##\hfil$\crcr#2\crcr\sim\crcr}}}
\makeatother
\input{epsf}
\begin{document}
\title{The interaction between H-dibaryons}
\author{T. Sakai, J. Mori$^{\ast}$, A.J. Buchmann$^{\ast\ast}$, 
K. Shimizu$^{\ast}$, K. Yazaki$^{\ast\ast\ast}$\\
{\em Research Center for Nuclear Physics, Osaka University}\\
$^{\ast}${\em Department of Physics, Sophia University}\\
$^{\ast\ast}${\em Institut f\"{u}r Theoretische Physik, Universit\"{a}t 
T\"{u}bingen}\\
$^{\ast\ast\ast}${\em Department of Physics, University of Tokyo}\\
}

\maketitle

\begin{abstract}
We investigate the interaction between H-dibaryons employing a quark 
cluster model with a one-gluon-exchange potential and an effective 
meson exchange potential (EMEP).
A deeply-bound state of two H-dibaryons due to the medium range 
attraction of the EMEP is obtained.
The bound H--H system has a size of about 0.8 $\sim$ 0.9 fm because of 
the short-range repulsion generated by the color-magnetic interaction 
and the Pauli principle.
\end{abstract}

\section{Introduction}

The H-dibaryon is a doubly strange six-quark state with spin and 
isospin 0.
Since its first prediction in 1977 \cite{Jaffe77}, 
it has been the subject of many theoretical and experimental 
studies as a possible candidate of a strongly bound exotic state. 
The main question is whether its mass is lighter than the 
$\Lambda\Lambda$ threshold, because such a light H-dibaryon is 
stable against strong decay.
However, theoretical estimates of its mass have varied widely 
ranging from deeply bound to unbound, depending on the
model or theory used for the calculations \cite{Href}.
Experimentally, there is no conclusive result yet, although 
several candidates for an H-dibaryon decay have been reported 
\cite{Shahbazian88}.
Many efforts to search for the H-dibaryon are under way \cite{Longacre95} 
and planned.
A double-hypernucleus event with a $\Lambda\Lambda$ binding energy 
less than about 30 MeV, has recently been found in an emulsion 
experiment at KEK \cite{Aoki91}. 
This excludes the possibility of a deeply-bound H-dibaryon.

The main dynamical factor responsible for a deeply-bound H-dibaryon 
is the color-magnetic interaction (CMI) in the one-gluon-exchange 
potential (OGEP).
The CMI plays an important role in explaining the low-lying 
hadron spectra and the short-range part of the baryon--baryon 
interaction.
In the H-dibaryon, the CMI is attractive.
On the other hand, it acts as a repulsive force for the nucleon-nucleon
interaction and also for almost all nucleon-hyperon and hyperon-hyperon 
interactions.
As is shown later, the CMI also generates a repulsive force between 
two H-dibaryons.
Another important factor determining the quark dynamics is the Pauli 
exclusion principle.
It is known that the Pauli principle gives repulsion for some channels 
of the nucleon--hyperon and hyperon--hyperon systems.  
Furthermore, because of the color exchange character of the OGEP, the CMI 
is only effective when the quark exchange interaction between different 
baryons is taken into account.

Quark exchanges between clusters can be correctly treated in the quark 
cluster model through the antisymmetrization of the wave function.
Therefore, we employ the quark cluster model to study the interaction
between H-dibaryons in this paper.

Assuming that the CMI plays a key role in determining the properties 
of the H--H interaction, Tamagaki suggested the possibility that a 
new form of hadronic matter, called H-matter, appears at densities 
several times higher than normal nuclear density \cite{Tamagaki91}.
In this pioneering study, a simple H--H interaction model consisting 
of a hard-core potential plus a square well attractive potential 
outside the core was used, because no microscopic calculation of the 
H--H interaction was available at that time.

In the present study, we extract more quantitative information on 
the H--H interaction by employing the quark cluster model used in 
Ref.\cite{Sakai92}.
In this model, quarks are confined by an inter-quark potential 
and subject to the OGEP \cite{Rujula75}.
The medium range attraction is taken into account by an effective 
meson exchange potential (EMEP) between H-dibaryons.
Here we take only flavour singlet scalar meson exchange between the 
H-dibaryons into account.
In the next section, we briefly explain the model used in the 
calculation and give the necessary matrix elements of the 
spin-flavor-color operators.
The results of the calculation and the discussion of the 
properties of the 
H--H interaction are given in section \ref{sec:results}.
Section \ref{sec:summary} is devoted to a summary.

\section{Model}

In the present quark cluster model, the H-dibaryon wave function 
$\phi(\xibm)$ is assumed to be a (0s)$^6$ configuration divided by 
the 0s wave function of the center-of-mass of the H-dibaryon with 
the size parameter $\sqrt{1/6}b$ for the orbital part, 
$\varphi (\xibm)$, where $\xibm$ is the internal coordinate.
\begin{equation}
\phi (\xibm) = \varphi (\xibm)
{\cal S} ([222]_{\rm SF}){\cal C}([222]_{\rm C}) .
\end{equation}
Here, ${\cal S} ([222]_{\rm SF})$ and ${\cal C}([222]_{\rm C})$ are 
the spin-flavor  and  color parts, respectively, with their 
permutational symmetries in parentheses.
The total wave function of the H--H system is written as
\begin{equation}
\Psi (\xibm _1, \xibm _2,\rbm)= {\cal A}[\phi (\xibm _1) \phi (\xibm _2) 
\chi (\rbm)] ,
\label{eq:rgmwf}
\end{equation}
where the wave function of the relative motion between the H-dibaryons 
$\chi(\rbm)$ is introduced and $\rbm$ is the relative coordinate 
between the H-dibaryons.
The wave function of the relative motion is further decomposed into 
components of the angular momentum of the relative motion, $L$, as
\begin{equation}
\chi (\rbm ) = \sum_{LM} \frac{\chi_L (r)}{r} Y_{LM} (\hat{\rbm} ) .
\label{eq:partwf}
\end{equation}
In this paper, only relative S-waves are considered. 

The antisymmetrization operator ${\cal A}$ can be expressed as
\begin{equation}
{\cal A} \equiv 1-36P_{67}+225P_{57}P_{68}-400P_{47}P_{58}P_{69} ,
\label{eq:asop}
\end{equation}
where $P_{ij}$ is the permutation operator.
Because each H-dibaryon consists of 6 quarks, there are 0,1,2 and 3 quark 
exchanges between H-dibaryons which are shown in Fig. \ref{f:norm}.
Because we symmetrize the two H--dibaryons system under a permutation of the
two H-dibaryons, 4,5 and 6 quark exchanges are taken into account by the 
0,1,2 and 3 quark exchanges and the exchange of the clusters as a whole. 

The Hamiltonian $H$ consists of the kinetic energy term, 
$K$, and the interaction term, $V$.
\begin{equation}
H=K+V,
\end{equation}
\begin{equation}
K=\displaystyle{\sum_{i=1}^{12} \frac{\pbm_i^2}{2m} } - K_{\rm G} ,
\end{equation}
where $K_{\rm G}$ is the center-of-mass energy of the twelve-quark system, 
and
\begin{equation}
V=\displaystyle{\sum_{i<j} } (V_{ij}^{\rm OGEP} + V_{ij}^{\rm conf}) .
\end{equation}
Here, $V_{ij}^{\rm OGEP}$ and $V_{ij}^{\rm conf}$ are the one-gluon-exchange 
potential (OGEP) and the confinement potential, respectively:
\begin{equation}
 V_{ij}^{\rm OGEP} = \frac{\alpha_{\rm s}}{4} \left[ \frac{1}{r_{ij}} - 
   \frac{\pi}{m_i m_j}\delta(r_{ij}) \left( 1+\frac{2}{3} \sigmabm_i \cdot 
   \sigmabm_j \right) \right] \lambdabm_i \cdot \lambdabm_j ,
\label{eq:OGEP}
\end{equation}
\begin{equation}
   V_{ij}^{\rm conf} = -a_{\rm c} r_{ij}^2 \lambdabm_i \cdot \lambdabm_j .
\end{equation}
The term containing $\sigmabm_i \cdot \sigmabm_j \lambdabm_i \cdot \lambdabm_j$
in eq.(\ref{eq:OGEP}) is known as the color-magnetic interaction (CMI) term. 
It is important for the short-range behavior of the baryon-baryon 
interaction and also plays a crucial role in the short-range H--H interaction.

Flavor SU(3) symmetry breaking in the CMI is taken into account by using the 
parametrization: 
$$\pi/m_i m_j \rightarrow \xi_{ij} \pi/m_{\rm u}^2,$$
where 
$\xi_{ij}=1$ when both $i$ and $j$ are not an s-quark, $\xi_{ij}=\xi_1$ when 
either $i$ or $j$ is an s-quark, and $\xi_{ij}=\xi_2$ when both $i$ and $j$ 
are s-quarks.
Note that we take $\xi_2$ as an independent parameter, instead of being 
$\xi_1^2$, which is straightforwardly expected from eq.(\ref{eq:OGEP}).
As in ref.\cite{Sakai92}, we use here the parameter set which reproduces 
the experimental values of the mass splittings among $\Lambda\Lambda$, 
N$\Xi$ and $\Sigma\Sigma$ correctly.

The other parameters entering OGEP are determined from the experimental
masses of octet and decuplet baryons \cite{Sakai92}.
Their values are reproduced here in Table \ref{tab1}.

The equation that determines the relative wave function $\chi(\rbm)$ is 
called the Resonating Group Method (RGM) equation.
\begin{equation}
\int \{ H_{\rm RGM} (\rbm, \rbm')-E N_{\rm RGM} (\rbm, \rbm') \} \chi (\rbm')
{\rm d}\rbm' =0, \label{eq:rgmeq}
\end{equation}
where $N_{\rm RGM}$ ($H_{\rm RGM}$) is the normalization and Hamiltonian 
kernel.
These kernels are calculated as matrix elements of $1$ and $H$ using 
the wave function of eq.(\ref{eq:rgmwf}). 

\begin{figure}[ht]
\caption{Ten types of two-body interaction with quark exchange}
\label{f:two}
\begin{picture}(400,160)
\put (20,80) {
\begin{picture}(50,70)
\put (1,65) {\line(0,-1){35}}
\put (6,65) {\line(0,-1){35}}
\put (1,47.5) {\line(1,0){5}}
\put (2,18)  {H}
\put (28,18) {H}
\put (25,0) {\makebox(0,0)[b]{Type 1}}
\end{picture}
}
\put (85,80) {
\begin{picture}(50,70)
\put (31,65) {\line(0,-1){35}}
\put (36,65) {\line(0,-1){35}}
\put (31,47.5) {\line(1,0){5}}
\put (2,18)  {H}
\put (28,18) {H}
\put (25,0) {\makebox(0,0)[b]{Type 2}}
\end{picture}
}
\put (175,80) {
\begin{picture}(50,70)
\put (6,65) {\line(0,-1){35}}
\put (26,65) {\line(0,-1){10}}
\put (26,55) {\line(-1,-1){15}}
\put (11,40) {\line(0,-1){10}}
\put (6,60) {\line(1,0){20}}
\put (2,18)  {H}
\put (28,18) {H}
\put (25,0) {\makebox(0,0)[b]{Type 3}}
\end{picture}
}
\put (265,80) {
\begin{picture}(50,70)
\put (11,65) {\line(0,-1){10}}
\put (11,55) {\line(1,-1){15}}
\put (26,40) {\line(0,-1){10}}
\put (31,65) {\line(0,-1){35}}
\put (11,60) {\line(1,0){20}}
\put (2,18)  {H}
\put (28,18) {H}
\put (25,0) {\makebox(0,0)[b]{Type 4}}
\end{picture}
}
\put (330,80) {
\begin{picture}(50,70)
\put (6,65) {\line(0,-1){35}}
\put (11,65) {\line(0,-1){10}}
\put (11,55) {\line(1,-1){15}}
\put (26,40) {\line(0,-1){10}}
\put (6,60) {\line(1,0){5}}
\put (2,18)  {H}
\put (28,18) {H}
\put (25,0) {\makebox(0,0)[b]{Type 5}}
\end{picture}
}
\put (20,0) {
\begin{picture}(50,70)
\put (26,65) {\line(0,-1){10}}
\put (26,55) {\line(-1,-1){15}}
\put (11,40) {\line(0,-1){10}}
\put (31,65) {\line(0,-1){35}}
\put (26,60) {\line(1,0){5}}
\put (2,18)  {H}
\put (28,18) {H}
\put (25,0) {\makebox(0,0)[b]{Type 6}}
\end{picture}
}
\put (85,0) {
\begin{picture}(50,70)
\put (11,65) {\line(0,-1){35}}
\put (26,65) {\line(0,-1){35}}
\put (11,60) {\line(1,0){15}}
\put (2,18)  {H}
\put (28,18) {H}
\put (25,0) {\makebox(0,0)[b]{Type 7}}
\end{picture}
}
\put (175,0) {
\begin{picture}(50,70)
\put (11,65) {\line(0,-1){10}}
\put (11,55) {\line(1,-1){15}}
\put (26,40) {\line(0,-1){10}}
\put (26,65) {\line(0,-1){10}}
\put (26,55) {\line(-1,-1){15}}
\put (11,40) {\line(0,-1){10}}
\put (11,60) {\line(1,0){15}}
\put (2,18)  {H}
\put (28,18) {H}
\put (25,0) {\makebox(0,0)[b]{Type 8}}
\end{picture}
}
\put (265,0) {
\begin{picture}(50,70)
\put (26,65) {\line(0,-1){10}}
\put (26,55) {\line(-4,-3){20}}
\put (6,40) {\line(0,-1){10}}
\put (31,65) {\line(0,-1){10}}
\put (31,55) {\line(-4,-3){20}}
\put (11,40) {\line(0,-1){10}}
\put (26,60) {\line(1,0){5}}
\put (2,18)  {H}
\put (28,18) {H}
\put (25,0) {\makebox(0,0)[b]{Type 9}}
\end{picture}
}
\put (330,0) {
\begin{picture}(50,70)
\put (6,65) {\line(0,-1){10}}
\put (6,55) {\line(4,-3){20}}
\put (26,40) {\line(0,-1){10}}
\put (11,65) {\line(0,-1){10}}
\put (11,55) {\line(4,-3){20}}
\put (31,40) {\line(0,-1){10}}
\put (6,60) {\line(1,0){5}}
\put (2,18)  {H}
\put (28,18) {H}
\put (25,0) {\makebox(0,0)[b]{Type 10}}
\end{picture}
}
\end{picture}
\end{figure}
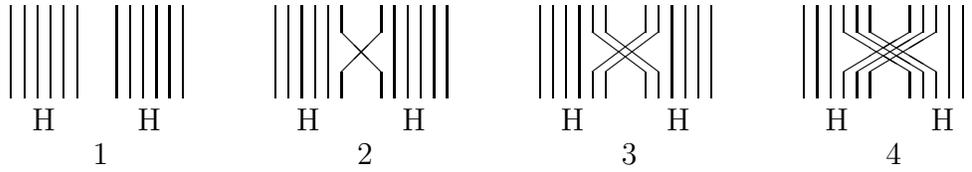

Details on how to evaluate the normalization and Hamiltonian kernels can 
be found, for example, in refs.~\cite{Oka81,Horiuchi77}. 
Here, we show the spin-flavor-color (SFC) parts of the matrix elements.
Because the SFC parts of the H-dibaryon wave functions are totally
antisymmetrized, it is sufficient to calculate the following matrix elements:
\begin{displaymath}
\langle H^{(\sigma\tau c)} 
H^{(\sigma\tau c)} \mid \xi_{ij} \lambdabm_i \cdot \lambdabm_j \mbox{(or }\xi_{ij} 
\lambdabm_i \cdot \lambdabm_j \sigmabm_i \cdot \sigmabm_j \mbox{)} \mid
H^{(\sigma\tau c)} H^{(\sigma\tau c)} \rangle 
\end{displaymath}
\begin{displaymath}
\langle H^{(\sigma\tau c)} H^{(\sigma\tau c)} \mid \xi_{ij} \lambdabm_i \cdot 
\lambdabm_j \mbox{(or }\xi_{ij} \lambdabm_i \cdot \lambdabm_j \sigmabm_i \cdot 
\sigmabm_j \mbox{)} P^{(\sigma\tau c)}_{67} \mid H^{(\sigma\tau c)} 
H^{(\sigma\tau c)} \rangle 
\end{displaymath}
\begin{displaymath}
\langle H^{(\sigma\tau c)} H^{(\sigma\tau c)} \mid \xi_{ij} \lambdabm_i \cdot 
\lambdabm_j \mbox{(or }\xi_{ij} \lambdabm_i \cdot \lambdabm_j \sigmabm_i \cdot 
\sigmabm_j \mbox{)} P^{(\sigma\tau c)}_{57} P^{(\sigma\tau c)}_{68} \mid
H^{(\sigma\tau c)} H^{(\sigma\tau c)} \rangle 
\end{displaymath}
\begin{equation}
\label{stcme}
\langle H^{(\sigma\tau c)} H^{(\sigma\tau c)} \mid \xi_{ij} \lambdabm_i \cdot 
\lambdabm_j \mbox{(or }\xi_{ij} \lambdabm_i \cdot \lambdabm_j \sigmabm_i \cdot 
\sigmabm_j \mbox{)} P^{(\sigma\tau c)}_{47} P^{(\sigma\tau c)}_{58} 
P^{(\sigma\tau c)}_{69} \mid H^{(\sigma\tau c)} H^{(\sigma\tau c)} \rangle .
\end{equation}
Here, $\mid H^{(\sigma\tau c)} H^{(\sigma\tau c)} \rangle $ is the SFC part of 
the shell model wave function in eq.~(\ref{eq:rgmwf}).
There are 66 pairs interacting via the two-body interaction, $V_{ij}$, 
and they are classified into ten topologically different types of graphs as 
illustrated in Fig. \ref{f:two}. 

To obtain, for a given number of quarks exchanged, the total contribution 
of each graph, we must multiply the matrix
element in eq.(\ref{stcme}) by the
number of ways of choosing the interacting pair without changing the structure 
of the graph and by the corresponding weight of the
quark exchange operators in eq.~(\ref{eq:asop}). After multiplying these 
factors, the matrix elements in SFC space can be written as 
\begin{equation}
A+B\xi_1+C\xi_2 .
\label{eq:abc}
\end{equation}
In Table \ref{t:ssll}, the values $A$, $B$ and $C$ for each type of 
graph and each number of quarks exchanged are shown together with 
the multiplying factors.

It is known that the OGEP is not sufficient to describe the medium-range part 
of the baryon--baryon interaction, and that an attractive contribution arising 
from the meson cloud is also necessary.
An effective meson exchange potential (EMEP) is introduced here as a force 
between H-dibaryons in order to simulate this meson contribution, although 
there are several works which introduce the meson exchange interactions as a 
force between quarks\cite{Fuji95,Straub88}.
The EMEP has previously been used to obtain a realistic description of 
the baryon--baryon interaction \cite{Oka81}.
We extend the EMEP to describe the flavor-independent scalar meson exchange 
interaction between H-dibaryons.
There are more elaborate meson exchange potentials \cite{Fuji95}, 
where other scalar and pseudoscalar meson exchanges are included,
in addition to the simple flavour-independent scalar meson exchange used here.
We shall argue later that the neglect of other meson exchanges will not 
substantially affect the qualitative conclusions obtained in this paper.

Here we introduce an EMEP which is assumed to be flavour-independent. 
Since the EMEP is expressed as a function of the relative coordinate
$\rbm$ between the H-dibaryons, we here explain how we add the EMEP to
the RGM equation in (\ref{eq:rgmeq}). 
First, we rewrite the RGM equation so that it has the same form as the usual 
Schr\"odinger equation, albeit with a non-local Hamiltonian.  
After introducing the renormalized H--H relative wave function, 
\begin{equation}
\tilde{\chi} ( \rbm ) = \int {\rm d} \rbm ' N_{\rm RGM}^{1/2} (\rbm , \rbm ') 
\chi ( \rbm ' ) , \label{eq:renowf}
\end{equation}
and the renormalized Hamiltonian, 
\begin{equation}
\tilde{H} (\rbm, \rbm') = \int {\rm d} \rbm '' {\rm d} \rbm ''' 
N_{\rm RGM}^{-1/2} (\rbm, \rbm'') [ H_{\rm RGM} (\rbm'', \rbm''') - E_{\rm int} 
N_{\rm RGM} (\rbm'', \rbm''') ] 
 N_{\rm RGM}^{-1/2} (\rbm''', \rbm') , 
\end{equation}
the RGM equation is equivalent to the following non-local Schr\"odinger 
equation:
\begin{equation}
\int {\rm d}\rbm' \tilde{H} (\rbm, \rbm') \tilde{\chi} (\rbm') = 
E_{\rm cm} \tilde{\chi} (\rbm) . 
\end{equation}
Here $E_{\rm int}$ is the internal energy, and $E_{\rm cm}$ is the energy 
of the relative motion in the center-of-mass system.  
Then we add the EMEP, ${\cal V} (r)$, to the renormalized 
Hamiltonian kernel $\tilde{H} (\rbm, \rbm')$, i.e., 
$\tilde{H} (\rbm, \rbm') \rightarrow \tilde{H} (\rbm, \rbm') + {\cal V} (r) 
 \delta (\rbm - \rbm')$.
For the EMEP, we take a gaussian-type function for simplicity, 
\begin{equation}
{\cal V} (r) = V_{\rm 0HH} \exp (-r^2/\alpha_{\rm HH}^2).
\label{EMEPfunc}
\end{equation}

The parameters of ${\cal V} (r)$, $\alpha_{\rm HH}$ and $V_{\rm 0HH}$, 
are obtained as follows.
In the same manner as in ref.\cite{Sakai92}, we employ the direct convolution 
of the EMEP determined from the scattering length and effective range 
for the ${}^1S_0$ state of the nucleon-nucleon interaction, and obtain
$\alpha_{\rm HH} = 1.00$ fm\footnote{The corresponding effective scalar meson 
mass $\mu_{\rm eff}$ can be obtained by equating the Fourier transform 
of the gaussian potential with that of the Yukawa-type for small momentum 
transfer, i.e., $\mu_{\rm eff} = 2/\alpha_{\rm HH} \simeq 400$ MeV, which is 
not very far from the empirical $\sigma$-meson mass.}.
The parameter $V_{\rm 0HH}$ is fixed so that the possible range of the 
H-dibaryon mass is consistent with the double hypernucleus event found at 
KEK \cite{Aoki91}, which is interpreted as $_{\Lambda\Lambda}^{10}$Be or 
$_{\Lambda\Lambda}^{13}$B with a $\Lambda\Lambda$ binding energy 
of $8.5 \pm 0.7$ or $27.6 \pm 0.7$ MeV, respectively.
The corresponding range obtained for $V_{\rm 0HH}$ is 
$-1096 \mbox{($\Lambda\Lambda$ 
threshold)} > V_{\rm 0HH} > -1227$ MeV ($_{\Lambda\Lambda}^{13}$B) [$-1136$ MeV 
($_{\Lambda\Lambda}^{10}$Be)].
On the other hand, $V_{\rm 0HH} = -1302$ MeV is obtained by the direct 
convolution of the EMEP determined from the nucleon--nucleon interaction, 
which gives a rather large 
binding energy of the H-dibaryon from the $\Lambda\Lambda$ threshold, 43.4 MeV.
The values obtained for $V_{\rm 0HH}$ are summarized in Table \ref{EMEP}.

\section{Numerical results and properties of the H--H interaction}
\label{sec:results}

In this section, we show the numerical results obtained by solving the RGM 
equation including the EMEP with the parameters explained in the previous
section.

\begin{figure}[htb]
\epsfysize=8cm
\epsfbox{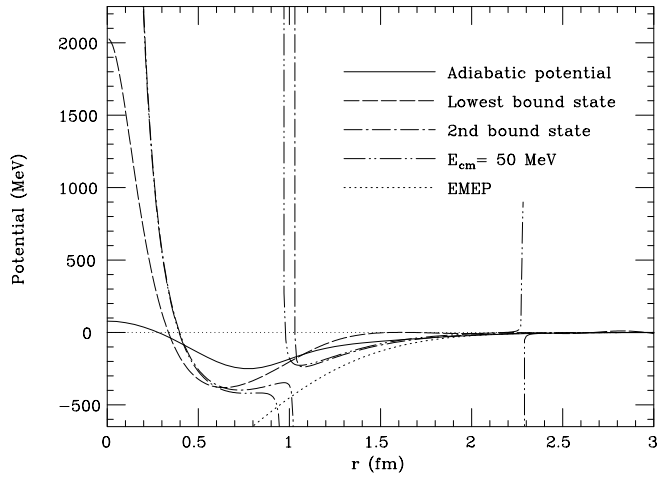}
\caption{The equivalent local potential (EQLP) for bound states
and a scattering state of the H--H system, 
and the adiabatic potential for $V_{\rm 0HH} = -1227$ MeV.
The solid curve denotes the adiabatic potential, the dashed 
curve denotes the EQLP of the lowest 
bound state, the chain curve denotes the EQLP of the second 
bound state, the double-dotted chain curve denotes the EQLP of 
the scattering state with center-of-mass energy $E_{\rm cm} = 50$ 
MeV.}
\label{fig:EQLP}
\end{figure}

In Fig.\ref{fig:EQLP}, the equivalent local potentials of bound and scattering 
states and the adiabatic potential for $V_{\rm 0HH} = -1227$ MeV are shown in 
order to appreciate the qualitative features of the H--H interaction, first.

The equivalent local potential ${\cal V}_{\rm EQLP}(r)$ \cite{Oka81} is 
defined as a potential which satisfies the following radial 
Schr\"{o}dinger equation:
\begin{equation}
-\frac{\hbar^2}{2\mu}\frac{{\rm d}^2}{{\rm d}r^2} \tilde{\chi}_0 (r) 
+ {\cal V}_{\rm EQLP} (r) \tilde{\chi}_0 (r) = E_{\rm cm} \tilde{\chi}_0 (r) ,
\label{eq:EQLP}
\end{equation}
where $\mu$ is the reduced mass and $\tilde{\chi}_0 (r)$ is the
renormalized wave function obtained by solving the RGM equation with the
non-local potential. The local potential obtained by this procedure
is called the trivially equivalent local potential.  We will discuss another
way of defining the equivalent local potential later.

The adiabatic potential $V_{\rm ad}(r)$ is often used in order to
get an idea in which regions of space repulsion and where attraction 
dominates. 
It is shown by the solid line in  Fig.\ref{fig:EQLP}.
The adiabatic potential is the ratio of the diagonal element 
of the GCM (Generator Coordinate Method: see refs.~\cite{Oka81,Horiuchi77}) 
Hamiltonian kernel to the GCM normalization kernel. 
\begin{equation}
V_{\rm ad} (r) = \frac{H_{\rm GCM} (r,r)}{N_{\rm GCM} (r,r)} - 
\left. \frac{H_{\rm GCM} (r,r)}{N_{\rm GCM} (r,r)} \right| _{r \rightarrow 
\infty} .
\end{equation}
Here the value at infinite H--H separation is subtracted, so that the 
$V_{\rm ad}(r)$  goes to zero as $r \rightarrow \infty$.
The GCM normalization and Hamiltonian kernels are the overlap and the 
expectation value of the Hamiltonian $H$ with a (0s$)^6$ configuration for
each H-dibaryon, where the origins of the two H-dibaryons are separated by
 a distance $r$.  Because
the origin is not the center-of-mass of the H-dibaryon, the adiabatic 
potential represents a smeared H--H potential, where the center-of-mass of
each H-dibaryon fluctuates.  Although the local and the non-local parts
of the potential are smeared out, 
the range of the fluctuations is sufficiently small to reflect the short-range
repulsion due to the effect of the 
exclusion principle and the CMI.

We see from Fig.\ref{fig:EQLP} that the CMI leads to a strong short-range 
repulsion. 
The short range part of the EQLP, which is due to the CMI has an 
energy-dependent repulsion, i.e., the second bound and scattering states 
feel a stronger repulsion at short distances than the ground state. 
This energy-dependence is the typical one also seen in 
the nucleon--nucleon potential \cite{Suzuki83}.  
This behavior is shown to be partly due to 
the non-locality of the potential \cite{Shimizu97}.
In the region below $r \lsim 2$ fm, the contribution from the OGEP has not 
yet died out and the  EQLP still differs from the EMEP 
in the medium-range although the EMEP dominates there. 
Furthermore, the energy dependence seems to be reversed compared to the one 
at short distances.  
Namely, at short distances, the low energy state has a weaker repulsion than 
the high energy state, whereas at long range, the repulsion for the low
 energy state becomes stronger  than the one for the high energy state.
If we employ the usual procedure \cite{Suzuki83} to obtain the energy-dependent
local potential from the non-local one, the ordering of the energy-dependence
would be the same for short distances and long distances.  

Our results for the energy-dependence of the H--H interaction can be understood 
in the following way.  
In the interacting region, the wave function $\tilde{\chi}_0 (r)$ 
for the attractive non-local potential is suppressed compared with
the wave function $\tilde{\chi}_{\rm EDL} (r)$ determined from the energy
dependent local (EDL) potential.
\[
\tilde{\chi}_0 (r)=\sqrt{\frac{m^{*}(r)}{m}}\tilde{\chi}_{\rm EDL} (r),
\]
where $m^{*}(r)$ is the $r$-dependent effective mass \cite{Negele83}.
This is known as the Perey-Buck effect \cite{Perey62}. 
In order to obtain the energy-dependent equivalent local potential,
we should use $\tilde{\chi}_{\rm EDL} (r)$.
In the actual calculation, however, we have employed
$\tilde{\chi}_0 (r)$ instead of $\tilde{\chi}_{\rm EDL} (r)$. Then the kinetic
energy given by the second derivative of the wave function 
$\tilde{\chi}_0 (r)$ differs from the one for $\tilde{\chi}_{\rm EDL} (r)$. 
This is the main reason for the different
energy dependences at short and long distances.

\begin{figure}[htb]
\epsfysize=8cm
\epsfbox{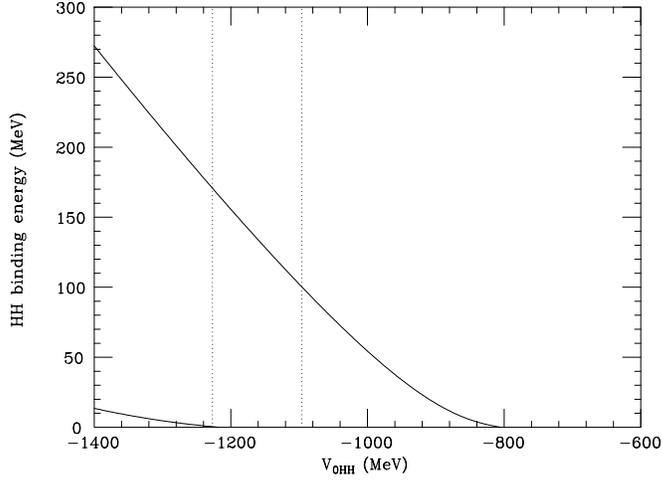}
\caption{The binding energy of the H--H system as a function of the strength 
parameter of the effective meson exchange potential $V_{\rm 0HH}$.
Both bound states are S-wave states.
The vertical dotted lines show the range of $V_{\rm 0HH}$ from $-1227$ to 
$-1096$ MeV.
}
\label{fig:BE}
\end{figure}

\begin{figure}[htb]
\epsfysize=8cm
\epsfbox{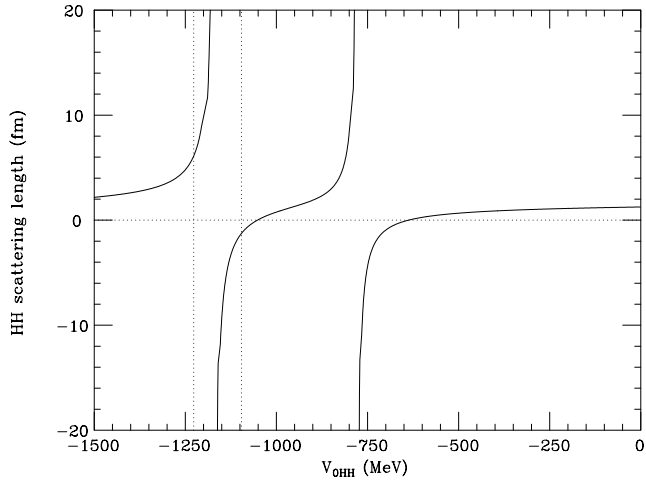}
\caption{The H--H scattering length $a$ as a function of the strength 
parameter of the effective meson exchange potential $V_{\rm 0HH}$. 
$a = -1.26$, $-4.91$, 6.12 and 3.51 fm for $V_{\rm 0HH} = -1096$, 
$-1136$, $-1227$ and $-1302$ MeV, respectively.
The vertical dotted lines show the range of $V_{\rm 0HH}$ from $-1227$ to 
$-1096$ MeV.
}
\label{fig:scattlen}
\end{figure}

\begin{figure}[htb]
\epsfysize=8cm
\epsfbox{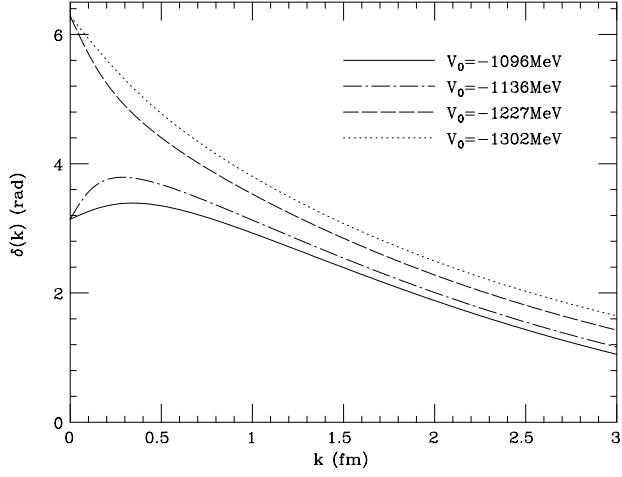}
\caption{The HH S-wave phase shift $\delta$ as a function of the relative 
wave number 
$k$ for several values of the strength parameter of the effective 
meson exchange potential $V_{\rm 0HH}$.}
\label{fig:phase}
\end{figure}

The binding energy of the H--H system as a function of the parameter 
$V_{\rm 0HH}$ is shown in Fig.\ref{fig:BE}.
The H--H system is a strongly self-binding system in the range,  
$-1096 > V_{\rm 0HH} > -1227$ MeV, and its binding energy in the ground state 
is 100$\sim$170 MeV.
Even a second bound state, which is also an S-wave state, appears in the range 
$V_{\rm 0HH} \lsim -1200$ MeV.
This is also seen from the divergent behavior of the H--H scattering length $a$ 
as a function of $V_{\rm 0HH}$ shown in Fig.\ref{fig:scattlen}, and from the
change of the S-wave phase shifts with $V_{\rm 0HH}$ shown in 
Fig.~\ref{fig:phase}.

\begin{figure}[htb]
\epsfysize=8cm
\epsfbox{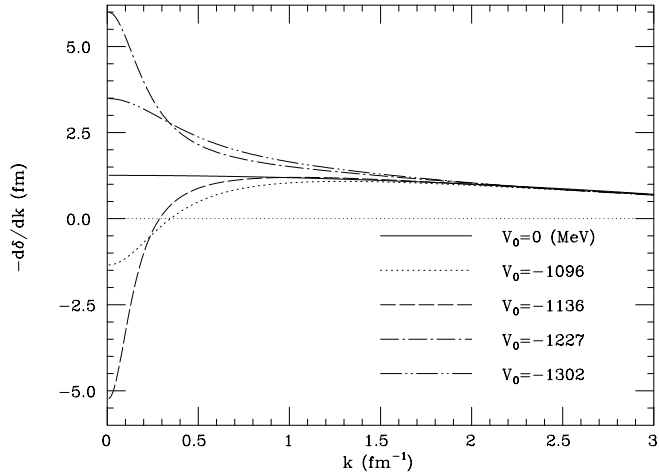}
\caption{The equivalent hard core radius as a function of the 
relative wave number 
$k$ for several values of the strength parameter of the effective 
meson exchange potential $V_{\rm 0HH}$.}
\label{fig:hardcore}
\end{figure}

Fig.\ref{fig:hardcore} shows the equivalent hard-core radius $a_{\rm h}$, 
which is defined as
\begin{equation}
a_{\rm h} = -\frac{{\rm d}\delta}{{\rm d}k} ,
\end{equation}
where $\delta$ is the phase shift, and 
$E_{\rm cm} = \hbar^2 k^2 / 2\mu$.
When $k$ goes to zero,  the value $a_{\rm h}$ coincides with 
the scattering length.
For comparison, we show $a_{\rm h}$ for the case without EMEP. 
For $k \gsim 2$ fm$^{-1}$, $a_{\rm h}$ is determined mainly 
from the repulsive core, and is of the order of 1 fm.

\begin{figure}[htb]
\epsfysize=8cm
\epsfbox{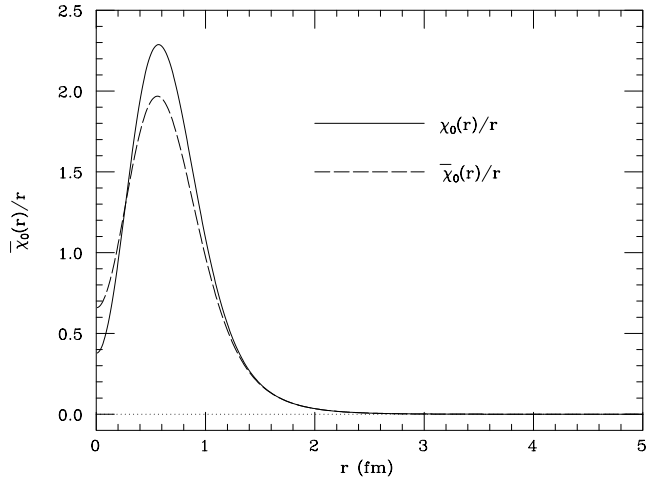}
\caption{The wave function of the lowest S-wave bound state for 
$V_{\rm 0HH} = -1227$ MeV.
The solid curve denotes the wave function without the renormalization 
$\chi_0 (r)/r$ and the dashed curve denotes the renormalized wave function 
$\tilde{\chi}_0 (r)/r$. The root-mean-square radii of the H--H system are 
0.92, 0.89, 0.84 and 0.81 fm for $V_{\rm 0HH} = -1096$, $-1136$, $-1227$ 
and $-1302$ MeV, respectively. 
}
\label{fig:wf}
\end{figure}

The wave function of the lowest bound state, $\tilde{\chi}_0(r)/r$, is shown 
for $V_{\rm 0HH} = -1227$ MeV in Fig.\ref{fig:wf}.
Because of the strong repulsive core, the peak of the wave function 
is around  0.7 fm in spite of the large binding energy of the H--H system.
The wave function is clearly localized around the minimum of the EQLP.
In order to see the effect of the renormalization, 
the unrenormalized wave function, $\chi_0(r)/r$, 
which does not contain the factor $N^{1/2}_{\rm RGM}$, 
is also shown.
The differences between the renormalized and unrenormalized wave functions 
can be understood as follows.
The square root of the normalization kernel is expanded 
in terms of the harmonic
oscillator wave functions, $u_{nlm}$, with the size 
parameter $\sqrt{2/3}b$ as
\begin{equation}
N^{1/2}_{\rm RGM} (\rbm , \rbm ') = \sum_{nlm} \sqrt{\mu_N} u_{nlm}^{\ast} ( 
\rbm ) u_{nlm} ( \rbm ') .
\label{eq:sqrtn}
\end{equation} 
Here, $\mu_N$ is the eigenvalue of the normalization kernel,
\begin{equation}
\mu_N = 1 -2 \left( \frac{2}{3} \right) ^N + \frac{49}{16} \left( \frac{1}{3} 
\right) ^N - \frac{41}{32} \delta_{N0} ,
\end{equation}
where $N = 2n+l$.
The angular momentum $l$ is zero for the S-wave interaction.
If $\mu_N = 1$ for all $N$ in the expansion of eq.(\ref{eq:sqrtn}), 
the left-hand side becomes $\delta$-function, $\delta (\rbm - \rbm')$.
The deviation of $\mu_N$ from 1 is due to the quark exchanges 
between the H-dibaryons.
The eigenvalues $\mu_N$ are $\frac{25}{32}$ for $n=0$ and 
$\frac{65}{144}$ for 
$n=1$, respectively.
The large contributions from $n=0$ 
and $n=1$ 
cause the enhancement of the amplitude 
around the origin and the suppression at intermediate range 
by the renormalization.

These results are obtained by employing the simple flavor-singlet EMEP.
Therefore, further examinations are required especially concerning the 
validity of the flavor independent EMEP.
Here, we wish to make a brief comment on the use of the EMEP which
is taken to simulate a flavor singlet scalar meson exchange.
Recently, more elaborate nucleon-hyperon interactions 
based on the SU(6) quark model \cite{Fuji95} have been suggested.  
In ref.~\cite{Fuji95}, flavor-octet scalar meson exchanges are also included, 
in addition to the flavor-singlet scalar meson.  
Although we do not take into account the octet mesons in the present 
calculation, we qualitatively discuss their effect on the H--H system. 
The potential due to the flavor-octet scalar-meson
exchange has the following structure:
\[
{\cal V}(r)=\fbm_1 \cdot \fbm_2 V(r),
\]
where $\fbm_1$ and $\fbm_2$ are flavor SU(3) operators for 
baryons 1 and 2, respectively.  This form is based on the SU(6) quark model.
In general, there are two types of meson-baryon couplings, 
which are called F and D.  
It is known that the F type coupling, which gives the above 
flavor-dependence for the baryon--baryon interaction, is dominant for
scalar meson exchanges \cite{Fuji95}. 
The expectation value of the
flavor operator for the H-dibaryon is given as follows.
\[
<H \mid \fbm_1 \cdot \fbm_2 \mid H>=\frac{1}{2}
\{ (\fbm_1+ \fbm_2)^2-\fbm_1^2-\fbm_2^2 \}=-12.
\]
Here, we have used that $<(\fbm_1+ \fbm_2)^2>=0$ for the H-dibaryon and 
that the H-dibaryon consists
of flavor-octet baryons for which $<\fbm_i^2>$ is 12.
Thus, the contributions from the flavor-octet scalar-meson exchanges 
in the H-dibaryon are repulsive because the radial part 
$V(r)$ is always attractive for scalar meson exchanges.  
As a result, if we include the flavor-octet mesons and fix the
strength of the EMEP from the binding energy of the H-dibaryon,
the strength of the flavor-singlet part becomes more 
attractive than in the present calculation without the flavor-octet mesons.  
For the H--H system, the flavor-octet meson exchanges do not
contribute because the H-dibaryon is a flavor SU(3) singlet.  
Therefore, if the flavor-octet mesons were included,
the H--H system would be more strongly bound. 

For the nucleon-nucleon system, octet scalar meson exchange 
is attractive in the ${}^1S_0$ channel. 
We expect, therefore, that the strength of the flavor-singlet part of the
EMEP derived from the nucleon--nucleon potential is decreased compared to the 
case without octet mesons. 
On the other hand, as discussed before, the strength of the 
flavor-singlet part of the EMEP as derived from the binding energy of the 
H-dibaryon is increased when the octet mesons are taken into account.  
Thus, the strengths derived from the nucleon--nucleon scattering data on 
one hand, and from the H-dibaryon binding energy on the other hand, 
would become similar when we include the octet scalar meson exchanges.
Recall that without the octet scalar mesons, the EMEP derived from
nucleon-nucleon scattering was more attractive ($-1302$ MeV) 
than the EMEP derived from the limits on the binding energy of the
H-particle ($-1227$ MeV $\sim $ $-1096$ MeV). 

Finally, we comment on the effect of pseudoscalar meson exchange.  
The pseudoscalar meson exchanges do not contribute to the H--H system.  
The contribution from pseudoscalar
meson exchanges to the binding energy of the H-dibaryon is known to be about 
5 MeV \cite{Koike90,Straub88}.
Therefore, even if we take into account the pseudoscalar meson exchanges,
there is no drastic change for the binding energy of the H--H system. 
Thus, the conclusion that the H--H system has a strong medium-range 
attraction leading to a bound state remains unchanged.

A bound state of two H-dibaryons is also obtained in the Skyrme 
model\cite{Issinskii88}.
The binding energy of the ``tetralambda" state is $E_{\rm B} = 15 \sim 20$ MeV, 
which is smaller than the value obtained in our calculation.

Finally, we comment on the implications of our results for the occurrence of 
H-matter.
Tamagaki \cite{Tamagaki91} discussed the possibility of a phase transition 
from neutron matter to H-matter at a density which is 6$\sim$9 times greater 
than the normal nuclear density $\rho_0$. In Tamagaki's calculation, the
H-dibaryons interact via a hard core potential and an attractive square 
well potential outside the core. 
In their work, the depth of the attractive potential was assumed to be so 
weak that it can be treated as a perturbation.
The depth of the square well potential has been determined as the value
where the scattering length becomes zero for the first time,
when the strength of the attractive potential is gradually increased, i.e.
as the limiting strength that still gives a positive scattering length. 
The pertinent strength parameter corresponding to our EMEP can 
be obtained from Fig.~\ref{fig:scattlen}. It is found to be 
$V_{\rm 0HH} = -638$ MeV. Thus the attraction used in Ref. ~\cite{Tamagaki91} 
is much weaker than the one used in the present calculation.
If the attractive H--H potential is indeed as strong as the EMEP 
employed in the present calculation, 
the critical transition density beyond which H-matter formation is 
energetically favorable may be appreciably lower. 
However, 
a quantitative estimate is not simple because of the inapplicability of 
perturbation theory under a strong attractive potential as in the present 
calculation.
A recalculation of the critical transition density for H-matter
formation is beyond the scope of the present work
(see however ref.\cite{Faessler97}).

\section{Summary}
\label{sec:summary}
We have investigated the interaction between two H-dibaryons in the quark 
cluster model with a one-gluon-exchange and quadratic confinement 
potential between  constituent quarks.
In addition, a phenomenological attractive meson exchange potential between 
H-dibaryons is used.
The parameters employed for the calculation are determined so as to reproduce 
the octet and decuplet baryon ground state masses and from the requirement 
that the H-dibaryon  mass is consistent with the observed binding energy of 
two $\Lambda$'s in a double hypernucleus.
These parameters were also used in our previous work on the N--H interaction
\cite{Sakai92,Sakai95}.
In our calculation, the effective scalar meson exchange potential is assumed 
to be a flavor SU(3) singlet.  
We have presented a qualitative discussion on the effects
of octet scalar mesons and pseudoscalar mesons.  
We conclude that the following properties of the H--H system remain unchanged 
even if we include the octet scalar and pseudoscalar mesons.

The main properties of the interaction between H-dibaryons can be 
characterized as 
a short-range repulsion due to Pauli blocking and the color-magnetic 
interaction, and a medium-range attraction due to 
flavor singlet scalar meson exchange.
As a result of the present RGM calculation, we have obtained a 
strongly-bound state of two H-dibaryons with a separation 
about 0.8 $\sim$ 0.9 fm due to the strong repulsive core.

The present results suggest that the critical density 
at which the formation of H-matter becomes energetically favorable
may be lower than the value $\rho_{\rm crit}=6 \rho_0$ 
found in Ref.~\cite{Tamagaki91}. 
This would have interesting astrophysical consequences. 
In the framework of the Walecka model, it has recently been shown 
\cite{Faessler97} that with the present H--H 
interaction, H-matter is unstable against compression. 
Thus, if the central density of a massive neutron star exceeds the critical
density for H-matter formation, the energetically favorable compression
of H-matter provides a possible scenario for the conversion of a neutron 
star into a strange quark star. 

\section*{Acknowledgements}

The work has been partly supported by
Grant-in-Aid for Scientific Research on Priority Areas (Theoretical Study
of Nuclei with Strangeness) by the Ministry of Education, Science, Sports
and Culture.
A. J. B. thanks the Japan Society for the Promotion of
Science (JSPS) for a postdoctoral fellowship during the years 1990-1991
when this work has been begun.

\clearpage

\begin{figure}[ht]
\caption{Quark exchange diagrams between two H--dibaryons} 
\label{f:norm}
\begin{picture}(400,80)
\put(20,0){
\begin{picture}(60,70)
\put (1,65) {\line(0,-1){35}}       
\put (6,65) {\line(0,-1){35}}
\put (11,65) {\line(0,-1){35}}
\put (16,65) {\line(0,-1){35}}
\put (21,65) {\line(0,-1){35}}
\put (26,65) {\line(0,-1){35}}
\put (41,65) {\line(0,-1){35}}
\put (46,65) {\line(0,-1){35}}
\put (51,65) {\line(0,-1){35}}
\put (56,65) {\line(0,-1){35}}
\put (61,65) {\line(0,-1){35}}
\put (66,65) {\line(0,-1){35}}
\put (9,18) {H}
\put (49,18){H}
\put (32,5) {1}
\end{picture}
}
\put(120,0){
\begin{picture}(60,70)
\put (1,65) {\line(0,-1){35}}
\put (6,65) {\line(0,-1){35}}
\put (11,65) {\line(0,-1){35}}
\put (16,65) {\line(0,-1){35}}
\put (21,65) {\line(0,-1){35}}
\put (26,65) {\line(0,-1){10}}
\put (26,40) {\line(0,-1){10}}
\put (41,65) {\line(0,-1){10}}
\put (41,40) {\line(0,-1){10}}
\put (46,65) {\line(0,-1){35}}
\put (51,65) {\line(0,-1){35}}
\put (26,55) {\line(1,-1){15}}
\put (26,40) {\line(1,1){15}}
\put (56,65) {\line(0,-1){35}}
\put (61,65) {\line(0,-1){35}}
\put (66,65) {\line(0,-1){35}}
\put (9,18) {H}
\put (49,18){H}
\put (32,5) {2}
\end{picture}
}
\put(220,0){
\begin{picture}(60,70)
\put (1,65) {\line(0,-1){35}}
\put (6,65) {\line(0,-1){35}}
\put (11,65) {\line(0,-1){35}}
\put (16,65) {\line(0,-1){35}}
\put (21,65) {\line(0,-1){10}}
\put (21,40) {\line(0,-1){10}}
\put (26,65) {\line(0,-1){10}}
\put (26,40) {\line(0,-1){10}}
\put (41,65) {\line(0,-1){10}}
\put (41,40) {\line(0,-1){10}}
\put (46,65) {\line(0,-1){10}}
\put (46,40) {\line(0,-1){10}}
\put (51,65) {\line(0,-1){35}}
\put (21,55) {\line(4,-3){20}}
\put (26,55) {\line(4,-3){20}}
\put (26,40) {\line(4,3){20}}
\put (21,40) {\line(4,3){20}}
\put (56,65) {\line(0,-1){35}}
\put (61,65) {\line(0,-1){35}}
\put (66,65) {\line(0,-1){35}}
\put (9,18) {H}
\put (49,18){H}
\put (32,5) {3}
\end{picture}
}
\put(320,0){
\begin{picture}(60,70)
\put (1,65) {\line(0,-1){35}}
\put (6,65) {\line(0,-1){35}}
\put (11,65) {\line(0,-1){35}}
\put (16,65) {\line(0,-1){10}}
\put (16,40) {\line(0,-1){10}}
\put (21,65) {\line(0,-1){10}}
\put (21,40) {\line(0,-1){10}}
\put (26,65) {\line(0,-1){10}}
\put (26,40) {\line(0,-1){10}}
\put (41,65) {\line(0,-1){10}}
\put (41,40) {\line(0,-1){10}}
\put (46,65) {\line(0,-1){10}}
\put (46,40) {\line(0,-1){10}}
\put (51,65) {\line(0,-1){10}}
\put (51,40) {\line(0,-1){10}}
\put (16,55) {\line(5,-3){25}}
\put (21,55) {\line(5,-3){25}}
\put (26,55) {\line(5,-3){25}}
\put (16,40) {\line(5,3){25}}
\put (21,40) {\line(5,3){25}}
\put (26,40) {\line(5,3){25}}
\put (56,65) {\line(0,-1){35}}
\put (61,65) {\line(0,-1){35}}
\put (66,65) {\line(0,-1){35}}
\put (9,18) {H}
\put (49,18){H}
\put (32,5) {4}
\end{picture}
}
\end{picture}
\end{figure}

\clearpage

\begin{table}
\caption{Quark model parameters}
\label{tab1}
\begin{center}
\begin{tabular}{cccccc}
\hline
$b$ (fm) & $m$ (MeV/$c^2$) & $\alpha_{\rm s}$ & $a_{\rm c}$ (MeV/fm$^2$) & 
$\xi_1$ & $\xi_2$ \\
\hline
0.6 & 300 & 1.394 & 33.0 & 0.6 & 0.1 \\
\hline
\end{tabular}
\end{center}
\end{table}

\renewcommand{\arraystretch}{0.8} 
\begin{footnotesize}
\begin{table}
\caption{The matrix elements of spin-flavor-color parts}
\label{t:ssll}
\begin{center}
\begin{tabular}{|c|c|cc||rrr|rrr|} \hline
number of exchanged quarks & overlap & \multicolumn{2}{c||}{interaction type} &
\multicolumn{3}{c|}{$\sigma_i\cdot\sigma_j\lambda_i\cdot\lambda_j$} &
\multicolumn{3}{c|}{$\lambda_i\cdot\lambda_j$} \\
(factor) & & \multicolumn{2}{c||}{(factor)} & $A$ & $B$ & $C$ & $A$ & $B$ & $C$ \\
 \hline
 & & $1$ & ( 15) & $5$ & $22$ & $-3$  & $-7$ & $-\frac{22}{3}$ & $-\frac{5}{3}$ \\  
 & & $2$ & (15) & $5$ & $22$ & $-3$ & $-7$ & $-\frac{22}{3}$ & $-\frac{5}{3}$ \\  
 & & $3$ & ( 0) & $0$ & $0$ & $0$ & $0$ & $0$ & $0$ \\   
 0  & & $4$ & ( 0) & $0$ & $0$ & $0$ & $0$ & $0$ & $0$ \\   
 & $1$ &  $5$ & ( 0) & $0$ & $0$  & $0$  & $0$  & $0$ & $0$ \\   
(1) & & $6$ & ( 0) & $0$ & $0$ & $0$ & $0$ & $0$ & $0$ \\   
 & & $7$ & (36) & $0$ & $0$ & $0$ & $0$ & $0$ & $0$ \\   
 & & $8$ & ( 0) & $0$ & $0$ & $0$ & $0$ & $0$ & $0$ \\   
 & & $9$ & ( 0) & $0$ & $0$ & $0$ & $0$ & $0$ & $0$ \\   
 & & $10$ & ( 0) & $0$ & $0$ & $0$ & $0$ & $0$ & $0$ \\ \hline 
 & & $1$ & (10) & $-\frac{20}{3}$ & $-\frac{88}{3}$ & $4$ & $\frac{28}{3}$ & $\frac{88}{9}$ & $\frac{20}{9}$ \\  
 & & $2$ & (10) & $-\frac{20}{3}$ & $-\frac{88}{3}$ & $4$ & $\frac{28}{3}$ & $\frac{88}{9}$  & $\frac{20}{9}$ \\  
 & & $3$ & ( 5) & $-\frac{10}{3}$ & $-\frac{44}{3}$ & $2$ & $\frac{14}{3}$ & $\frac{44}{9}$ & $\frac{10}{9}$ \\   
 1 & & $4$ & ( 5) & $-\frac{10}{3}$ & $-\frac{44}{3}$ & $2$ & $\frac{14}{3}$ & $\frac{44}{9}$ & $\frac{10}{9}$ \\   
 & $-2$ &  $5$ & ( 5) & $-\frac{10}{3}$ & $-\frac{44}{3}$ & $2$ & $\frac{14}{3}$ & $\frac{44}{9}$ & $\frac{10}{9}$ \\    
(36) & & $6$ & ( 5) & $-\frac{10}{3}$ & $-\frac{44}{3}$ & $2$ & $\frac{14}{3}$ & $\frac{44}{9}$  & $\frac{10}{9}$ \\   
 & & $7$ & (25) & $-\frac{89}{16}$ & $\frac{11}{24}$ & $-\frac{139}{48}$ & $-\frac{683}{144}$ & $-\frac{341}{72}$ & $-\frac{19}{16}$ \\   
 & & $8$ & ( 1) & $-\frac{64}{3}$ & $0$ & $-\frac{32}{3}$ & $-\frac{64}{9}$ & $0$ & $-\frac{32}{9}$ \\   
 & & $9$ & ( 0) & $0$ & $0$ & $0$ & $0$ & $0$ & $0$ \\   
 & & $10$ & ( 0) & $0$ & $0$ & $0$ & $0$ & $0$ & $0$ \\ \hline 
 & & $1$ & ( 6) & $\frac{13073}{1296}$ & $\frac{2335}{72}$ & $-\frac{3971}{1296}$ & $-\frac{11851}{1296}$ & $-\frac{21623}{1944}$ & $-\frac{6965}{3888}$ \\  
 & & $2$ & ( 6) & $\frac{13073}{1296}$ & $\frac{2335}{72}$ & $-\frac{3971}{1296}$ & $-\frac{11851}{1296}$ & $-\frac{21623}{1944}$ & $-\frac{6965}{3888}$ \\  
 & & $3$ & ( 8) & $-\frac{7}{8}$ & $\frac{253}{12}$ & $-\frac{137}{24}$ & $-\frac{233}{24}$ & $-\frac{253}{36}$ & $-\frac{223}{72}$ \\   
 2 & & $4$ & ( 8) & $-\frac{7}{8}$ & $\frac{253}{12}$ & $-\frac{137}{24}$ & $-\frac{233}{24}$ & $-\frac{253}{36}$ & $-\frac{223}{72}$ \\   
 & $\frac{49}{16}$ & $5$ & ( 8) & $-\frac{7}{8}$ & $\frac{253}{12}$ & $-\frac{137}{24}$ & $-\frac{233}{24}$ & $-\frac{253}{36}$ & $-\frac{223}{72}$ \\   
(225)& & $6$ & ( 8) & $-\frac{7}{8}$ & $\frac{253}{12}$ & $-\frac{137}{24}$ & $-\frac{233}{24}$ & $-\frac{253}{36}$ & $-\frac{223}{72}$ \\   
 & & $7$ & (16) & $\frac{36281}{3888}$ & $-\frac{6593}{1944}$ & $\frac{21437}{3888}$ & $\frac{18347}{1944}$ & $\frac{7357}{972}$ & $\frac{5495}{1944}$ \\   
 & & $8$ & ( 4) & $\frac{8227}{243}$ & $\frac{1786}{243}$ & $\frac{3667}{243}$ & $\frac{23611}{1944}$ & $\frac{2093}{972}$ & $\frac{10759}{1944}$ \\   
 & & $9$ & ( 1) & $\frac{3953}{648}$ & $\frac{499}{36}$ & $-\frac{269}{648}$ & $-\frac{1675}{648}$ & $-\frac{4187}{972}$ & $-\frac{419}{1944}$ \\   
 & & $10$ & ( 1) & $\frac{3953}{648}$ & $\frac{499}{36}$ & $-\frac{269}{648}$ & $-\frac{1675}{648}$ & $-\frac{4187}{972}$ & $-\frac{419}{1944}$ \\ \hline  
 & & $1$ & ( 3) & $-\frac{1613}{576}$ & $-\frac{6545}{864}$ & $\frac{853}{1728}$ & $\frac{3881}{1728}$ & $\frac{2527}{864}$ & $\frac{677}{1728}$ \\  
 & & $2$ & ( 3) & $-\frac{1613}{576}$ & $-\frac{6545}{864}$ & $\frac{853}{1728}$ & $\frac{3881}{1728}$ & $\frac{2527}{864}$ & $\frac{677}{1728}$ \\  
 & & $3$ & ( 9) & $\frac{3}{32}$ & $-\frac{171}{16}$ & $\frac{87}{32}$ & $\frac{143}{32}$ & $\frac{57}{16}$ & $\frac{43}{32}$ \\   
 3 & & $4$ & ( 9) & $\frac{3}{32}$ & $-\frac{171}{16}$ & $\frac{87}{32}$ & $\frac{143}{32}$ & $\frac{57}{16}$ & $\frac{43}{32}$ \\   
 & $-\frac{41}{32}$ & $5$ & ( 9) & $\frac{3}{32}$ & $-\frac{171}{16}$ & $\frac{87}{32}$ & $\frac{143}{32}$ & $\frac{57}{16}$ & $\frac{43}{32}$ \\   
(400)& & $6$ & ( 9) & $\frac{3}{32}$ & $-\frac{171}{16}$ & $\frac{87}{32}$ & $\frac{143}{32}$ & $\frac{57}{16}$ & $\frac{43}{32}$ \\   
 & & $7$ & ( 9) & $-\frac{33185}{2592}$ & $-\frac{5239}{1296}$ & $-\frac{13973}{2592}$ & $-\frac{13611}{2592}$ & $-\frac{863}{432}$ & $-\frac{1837}{864}$ \\   
 & & $8$ & ( 9) & $-\frac{31319}{1296}$ & $-\frac{5845}{648}$ & $-\frac{12737}{1296}$ & $-\frac{8031}{1296}$ & $-\frac{23}{216}$ & $-\frac{1327}{432}$ \\   
 & & $9$ & ( 3) & $-\frac{2131}{576}$ & $-\frac{8575}{864}$ & $\frac{1091}{1728}$ & $\frac{3895}{1728}$ & $\frac{2513}{864}$ & $\frac{691}{1728}$ \\   
 & & $10$ & ( 3) & $-\frac{2131}{576}$ & $-\frac{8575}{864}$ & $\frac{1091}{1728}$ & $\frac{3895}{1728}$ & $\frac{2513}{864}$ & $\frac{691}{1728}$ \\    \hline 
\end{tabular}   
\end{center}
\end{table}
\end{footnotesize}
\thispagestyle{empty}

\begin{table}
\caption{Strength parameter of the effective meson exchange potential}
\label{EMEP}
\begin{center}
\begin{tabular}{ccl}
\hline
$V_{\rm 0HH}$ (MeV) & $2M_{\Lambda}-M_{\rm H}$ (MeV) & \hspace{1mm} \\
\hline
$-1096$ &    0 & $\Lambda\Lambda$ threshold \\
$-1136$ &  8.5 & $^{10}_{\Lambda\Lambda}$Be \\
$-1227$ & 27.6 & $^{13}_{\Lambda\Lambda}$B \\
$-1302$ & 43.4 & \hspace{1mm} \\
\hline
\end{tabular}
\end{center}
\end{table}

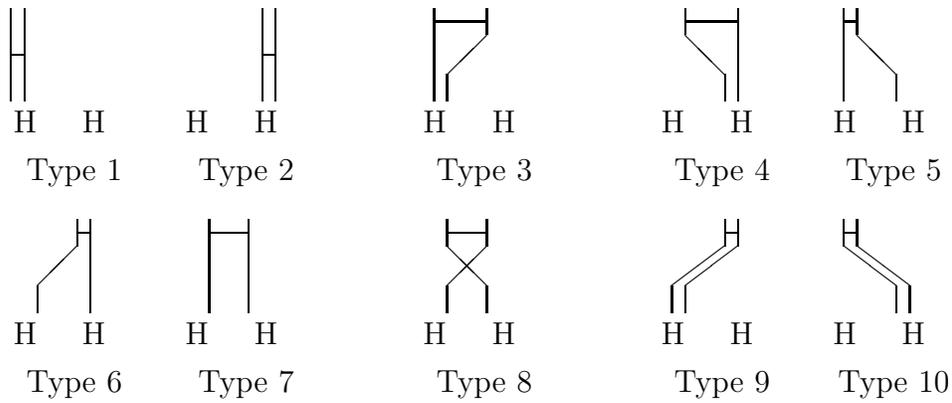
\end{document}